\documentstyle[aps,draft]{revtex}

\begin{document}
\draft
\title{Momentum distribution of confined bosons: temperature dependence}
\author{J.\ Tempere$^{1}$, F. Brosens$^{1}$, L. F. Lemmens$^{2}$, J. T. Devreese$%
^{1,2,}$\thanks{%
Also at Technische Universiteit Eindhoven, NL 5600 MB Eindhoven, The
Netherlands.}}
\date{Accepted in Phys. Rev. A on July 1, 1998}
\address{$^1$ Universiteit Antwerpen (UIA), Universiteitsplein 1, B2610 Antwerpen,\\
Belgium.\\
$^2$ Universiteit Antwerpen (RUCA), Groenenborgerlaan 171, B2020 Antwerpen,\\
Belgium.}
\maketitle

\begin{abstract}
The momentum distribution function of a parabolically confined gas of bosons
with harmonic interparticle interactions is derived. In the Bose-Einstein
condensation region, this momentum distribution substantially deviates from
a Maxwell-Boltzmann distribution. It is argued that the determination of the
temperature of the boson gas from the Bose-Einstein momentum distribution
function is more appropriate than the currently used fitting to the high
momentum tail of the Maxwell-Boltzmann distribution.
\end{abstract}


\section{Introduction}

The extreme low temperature required for the observation of Bose-Einstein
condensation alkali vapors is a major achievement of present day
experimentation \cite{jila,mit,BraPRL95}. Unfortunately, the temperature
itself is not directly observable. It has to be estimated from other, more
accessible properties of the system, such as the velocity distribution of
the vapor \cite{jila2}. Because of the uncertainty relation between the
inverse temperature and the energy of the system for a finite number of
particles, it is not clear {\sl a priori} how many particles are required
for a reasonable thermodynamically defined concept of temperature. In the
present paper, we do not address this fundamental question, but rather
assume that the number of bosons in the system is sufficiently large to have
a meaningful definition of the temperature.

Under this assumption, the experimental estimate of the temperature\cite
{jila2} is based on the hypothesis that the momentum distribution function
is Maxwell-Boltzmann like, at least for sufficiently high momentum. This
hypothesis deserves further investigation and quantitative comparison to the
momentum distribution of a parabolically confined Bose-Einstein gas. To the
best of our knowledge, the Bose-Einstein momentum distribution in a
parabolic confinement potential has not been documented before.

Using the methods for path integration developed in \cite
{LBDpre96,BDLpre97a,BDLpre97b}, we have been able to calculate the momentum
distribution function of a gas of parabolically confined bosons with
quadratic harmonic interparticle interactions. This calculation is presented
in Sec. II. Other investigations using the same methodology \cite{TBDLnyp97}
indicated that for weakly interacting systems, such as $^{87}$Rb and $^{31}$%
Na, the interaction slightly influences the parameters involved in the model
system, and therefore have only a moderate influence the momentum
distribution.

In section III, the approximations are examined which lead to the Maxwellian
momentum distribution function, and a quantitative comparison is made
between the temperature obtained in the Bose-Einstein distribution and the
estimate of the temperature, obtained by fitting a Maxwell-Boltzmann
distribution to the tail of the measured velocity distribution. The
conclusion from this analysis is that the estimate from the Maxwellian
deviates substantially from the temperature of the Bose-Einstein
distribution in the presence of a Bose-Einstein condensate.

\section{Momentum distribution function of parabolically trapped bosons}

The anisotropic one-body potential energy $V_1$ and the two-body potential
energy $V_2$ of the model system consisting of $N$ bosons are given by 
\begin{equation}
V_1=\frac m2%
\mathop{\displaystyle\sum}%
\limits_{j=1}^N\left( \Omega _{xy}^2(x_j^2+y_j^2)+\Omega _z^2z_j^2\right) 
\text{ and }V_2=\mp \frac{m\omega ^2}4%
\mathop{\displaystyle\sum}%
\limits_{j,l=1}^N\left( {\bf r}_j-{\bf r}_l\right) ^2
\end{equation}
where ${\bf r}_j=(x_j,y_j,z_j)$ is the position of the $j$th boson, $m$ is
the mass of the bosons and $\Omega _{xy}$, $\Omega _z$ are the frequencies
of the parabolic confinement in the $xy$-plane and along the $z$-axis,
respectively. The model interparticle interaction is harmonic with frequency 
$\omega $ and can be either attractive (--) or repulsive (+). The
renormalized frequencies in the $xy$-plane and along the $z$-axis will be
denoted by $w_{xy}=\sqrt{\Omega _{xy}^2\mp N\omega ^2}$ and $w_z=\sqrt{%
\Omega _z^2\mp N\omega ^2}$, respectively. These frequencies are the
eigenfrequencies of the internal degrees of freedom of the boson gas,
whereas the confining frequencies $\Omega _{xy}$, $\Omega _z$ are the
eigenfrequencies for the center-of-mass motion. The momentum distribution $%
n_{\text{bose}}({\bf p})$, giving the statistical mean number of bosons with
a given momentum ${\bf p}$ in the canonical ensemble, can be written as 
\begin{eqnarray}
n_{\text{bose}}({\bf p}) &=&\left\langle 
\mathop{\displaystyle\sum}%
\limits_{j=1}^N\delta ({\bf p}-{\bf p}_j)\right\rangle  \label{veld} \\
&=&\int \frac{d{\bf q}}{(2\pi )^3}\text{ }e^{i{\bf q\cdot p}}\left\langle 
\mathop{\displaystyle\sum}%
\limits_{j=1}^Ne^{-i{\bf q\cdot p}_j}\right\rangle ,
\end{eqnarray}
where the angular brackets indicate quantum averages for boson statistics.
These mean values can be calculated for a system at temperature $T=1/\left(
k_B\beta \right) $ using the path integral propagator in momentum space $%
{\cal K}({\bf p}_1^{\prime \prime },..,{\bf p}_N^{\prime \prime };\beta |%
{\bf p}_1^{\prime },..,{\bf p}_N^{\prime };0)$: 
\begin{equation}
\left\langle A({\bf p}_1,..,{\bf p}_N)\right\rangle =\frac{\left( 
\mathop{\displaystyle\prod}%
\limits_{j=1}^N\int d{\bf p}_j\right) A({\bf p}_1,..,{\bf p}_N){\cal K}({\bf %
p}_1,..,{\bf p}_N;\beta |{\bf p}_1,..,{\bf p}_N;0)}{\left( 
\mathop{\displaystyle\prod}%
\limits_{j=1}^N\int d{\bf p}_j\right) {\cal K}({\bf p}_1,..,{\bf p}_N;\beta |%
{\bf p}_1,..,{\bf p}_N;0)}.  \label{meanval}
\end{equation}

The calculation of the momentum distribution, analogous to the calculation
of the spatial density distribution $n({\bf r})$ in \cite{BDLpre97b}, can be
summarized as follows. First the propagator in momentum space is evaluated
by a Fourier transform of the corresponding propagator in position space
obtained in \cite{BDLpre97b}. This results in

\begin{equation}
{\cal K}({\bf p}_1,..,{\bf p}_N;\beta |{\bf p}_1,..,{\bf p}_N;0)=\frac 1{N!}%
\mathop{\displaystyle\sum}%
\limits_{\text{permutations }P}{\cal K}_{\text{d}}({\bf p}_1,..,{\bf p}%
_N;\beta |P{\bf p}_1,..,P{\bf p}_N;0),  \label{pm1}
\end{equation}
with 
\begin{eqnarray}
{\cal K}_{\text{d}}({\bf p}_1,..,{\bf p}_N;\beta |P{\bf p}_1,..,P{\bf p}%
_N;0) &=&\sqrt{%
{\displaystyle{w_z\sinh \left( \beta \hbar w_z\right)  \over \Omega _z\sinh \left( \beta \hbar \Omega _z\right) }}%
{{\frac{w_{xy}^2\sinh ^2\left( \beta \hbar w_{xy}\right) }{\Omega
_{xy}^2\sinh ^2\left( \beta \hbar \Omega _{xy}\right) }}}}%
\mathop{\displaystyle\prod}%
\limits_{j=1}^NK_{\text{d}}({\bf p}_j;\beta |P{\bf p}_j;0)  \nonumber \\
&&\times \exp \left\{ 
\begin{array}{c}
-%
{\displaystyle{P_z^2 \over \hbar mN}}%
\left( 
{\displaystyle{\tanh \left( \beta \hbar \Omega _z/2\right)  \over \Omega _z}}%
-%
{\displaystyle{\tanh \left( \beta \hbar w_z/2\right)  \over w_z}}%
\right) \\ 
-%
{\displaystyle{P_x^2+P_y^2 \over \hbar mN}}%
\left( 
{\displaystyle{\tanh \left( \beta \hbar \Omega _{xy}/2\right)  \over \Omega _{xy}}}%
-%
{\displaystyle{\tanh \left( \beta \hbar w_{xy}/2\right)  \over w_{xy}}}%
\right)
\end{array}
\right\} ,  \label{pm2}
\end{eqnarray}
where ${\bf P}=\sum_{j=1}^N{\bf p}_j$ is the total momentum, and $K_{\text{d}%
}({\bf p}^{\prime \prime };\beta |{\bf p}^{\prime };0)$ is the one-particle
propagator in momentum space for a particle in an anisotropic harmonic
potential. Next, the sum over permutations appearing in the expression (\ref
{meanval}) for the mean value $\left\langle 
\mathop{\textstyle\sum}%
\nolimits_{j=1}^Ne^{-i{\bf q\cdot p}_j}\right\rangle $ is transformed into a
cyclic sum (with $M_\ell $ the number of cycles of length $\ell $). The
cyclic summation can be performed for the generating function 
\begin{equation}
g_1(u)=\sum_{N=1}^\infty \left\langle 
\mathop{\displaystyle\sum}%
\limits_{j=1}^Ne^{-i{\bf q\cdot p}_j}\right\rangle u^NZ(N),
\end{equation}
where $Z(N)=\left( \prod_{j=1}^N\int d{\bf p}_j\right) {\cal K}({\bf p}_1,..,%
{\bf p}_N;\beta |{\bf p}_1,..,{\bf p}_N;0)$ is the partition sum. The result
of the cyclic summation is 
\begin{eqnarray}
g_1(u) &=&\left( \sum_{n=1}^\infty u^nZ(n)\right) \exp \left\{ 
\begin{array}{c}
-%
{\displaystyle{\hbar mq_z^2 \over 4N}}%
\left( 
{\displaystyle{\Omega _z \over \coth (\beta \hbar \Omega _z/2)}}%
-%
{\displaystyle{w_z \over \coth (\beta \hbar w_z/2)}}%
\right) \\ 
-%
{\displaystyle{\hbar m(q_x^2+q_y^2) \over 4N}}%
\left( 
{\displaystyle{\Omega _{xy} \over \coth (\beta \hbar \Omega _{xy}/2)}}%
-%
{\displaystyle{w_{xy} \over \coth (\beta \hbar w_{xy}/2)}}%
\right)
\end{array}
\right\}  \nonumber \\
&&\times 
\mathop{\displaystyle\sum}%
\limits_{\ell =1}^\infty \frac{u^\ell \exp \left\{ -\frac{\hbar m}4\left[
q_z^2w_z\coth (\ell \beta \hbar w_z/2)+(q_x^2+q_x^2)w_z\coth (\ell \beta
\hbar w_z/2)\right] \right\} }{8\sinh \left( \ell \beta \hbar w_z/2\right)
\sinh ^2\left( \ell \beta \hbar w_{xy}/2\right) }.
\end{eqnarray}
The expectation value $\left\langle 
\mathop{\textstyle\sum}%
\nolimits_{j=1}^Ne^{-i{\bf q\cdot p}_j}\right\rangle $ for $N$ bosons can
then be derived in closed form from the generating functional $g_1(u)$ by
collecting all the terms with the same power in $u$: 
\begin{eqnarray}
\left\langle 
\mathop{\displaystyle\sum}%
\limits_{j=1}^Ne^{-i{\bf q\cdot p}_j}\right\rangle &=&\exp \left\{ 
\begin{array}{c}
-%
{\displaystyle{\hbar mq_z^2 \over 4N}}%
\left( 
{\displaystyle{\Omega _z \over \coth (\beta \hbar \Omega _z/2)}}%
-%
{\displaystyle{w_z \over \coth (\beta \hbar w_z/2)}}%
\right) \\ 
-%
{\displaystyle{\hbar m(q_x^2+q_y^2) \over 4N}}%
\left( 
{\displaystyle{\Omega _{xy} \over \coth (\beta \hbar \Omega _{xy}/2)}}%
-%
{\displaystyle{w_{xy} \over \coth (\beta \hbar w_{xy}/2)}}%
\right)
\end{array}
\right\}  \nonumber \\
&&\times 
\mathop{\displaystyle\sum}%
\limits_{\ell =1}^N\frac{Z(N-\ell )}{Z(N)}\frac{\exp \left\{ -\frac{\hbar m}4%
\left[ 
\begin{array}{c}
q_z^2w_z\coth (\ell \beta \hbar w_z/2) \\ 
+(q_x^2+q_x^2)w_z\coth (\ell \beta \hbar w_z/2)
\end{array}
\right] \right\} }{8\sinh \left( \ell \beta \hbar w_z/2\right) \sinh
^2\left( \ell \beta \hbar w_{xy}/2\right) }.  \label{expval}
\end{eqnarray}
Finally, the momentum distribution (\ref{veld}) is found as a Fourier
transform of the expectation value (\ref{expval}). The result is: 
\begin{eqnarray}
n_{\text{bose}}({\bf p}) &=&\left( \frac 1{4\pi \hbar m}\right) ^{3/2}%
\mathop{\displaystyle\sum}%
\limits_{\ell =1}^N%
{\displaystyle{Z(N-\ell ) \over Z(N)\sinh ^2\left( \frac{\beta \hbar \Omega _{xy}\ell }2\right) \sinh \left( \frac{\beta \hbar \Omega _z\ell }2\right) }}%
\nonumber \\
&&\times \frac{\exp \left( -%
{\displaystyle{(p_x^2+p_y^2) \over \hbar mA_\ell (\Omega _{xy},w_{xy})}}%
-%
{\displaystyle{p_z^2 \over \hbar mA_\ell (\Omega _z,w_z)}}%
\right) }{A_\ell \left( \Omega _{xy},w_{xy}\right) \sqrt{A_\ell (\Omega
_z,w_z)}},  \label{exact}
\end{eqnarray}
with 
\begin{equation}
A_\ell \left( \Omega ,w\right) =\frac w{\tanh (\beta \hbar w\ell /2)}+\frac 1%
N\left[ \frac \Omega {\tanh (\beta \hbar \Omega /2)}-\frac w{\tanh (\beta
\hbar w/2)}\right] .
\end{equation}
In figure 1 this momentum distribution is shown at different temperatures
for an isotropically confined Bose gas with $w=\Omega $ and consisting of
1000 bosons. The momentum distribution for $w/\Omega \neq 1$ is
qualitatively similar to the distribution with $w/\Omega =1$ shown in figure
1. In the inset, the value of the momentum distribution in the origin is
shown as a function of the temperature. As the temperature is lowered below
the condensation temperature, the momentum distribution becomes more
pronounced in the origin: the average number of bosons with momentum zero
rises sharply.

\section{Approximate velocity distributions}

The momentum distribution (\ref{exact}) calculated in the previous section
for a gas of parabolically confined bosons can be related to the Maxwell
distribution through a series of approximations. First, in the case of a
non-interacting ($\omega \rightarrow 0$), parabolically confined boson gas
the momentum distribution (\ref{exact}) becomes 
\begin{equation}
n_{\text{bose}}^{\text{ideal}}\left( {\bf p}\right) =\left( \frac 1{4\pi
\hbar m}\right) ^{3/2}%
\mathop{\displaystyle\sum}%
\limits_{\ell =1}^N 
{\displaystyle{Z(N-\ell )\exp \left\{ %
\begin{array}{c}
-%
{\displaystyle{(p_x^2+p_y^2) \over \hbar m\Omega _{xy}\coth (\beta \hbar \Omega _{xy}\ell /2)}} \\ 
-%
{\displaystyle{p_z^2 \over \hbar m\Omega _z\coth (\beta \hbar \Omega _z\ell /2)}}%
\end{array}
\right\}  \over Z(N)\Omega _{xy}\sinh \left( \beta \hbar \Omega _{xy}\ell \right) \sqrt{\Omega _z\sinh \left( \beta \hbar \Omega _z\ell \right) }}}%
.  \label{vid}
\end{equation}
Furthermore, according to Feynman \cite{Feyn1}, the Maxwell-Boltzman
statistics can be recovered by considering only the cycles of length $\ell
=1 $ in the cyclic summations of (\ref{exact}). This means that permutations
involving cycles of length larger than 1 play no role any more. Hence the
momentum distribution function for distinguishable non-interacting particles
in a parabolic potential is: 
\begin{equation}
n_{\text{dist.}}^{\text{ideal}}\left( {\bf p}\right) =\left( 
{\displaystyle{\tanh ^2(\beta \hbar \Omega _{xy}/2)\tanh (\beta \hbar \Omega _z/2) \over (\pi \hbar m)^3\Omega _{xy}^2\Omega _z}}%
\right) ^{1/2}\exp \left\{ 
\begin{array}{c}
- 
{\displaystyle{(p_x^2+p_y^2) \over \hbar m\Omega _{xy}\coth (\beta \hbar \Omega _{xy}/2)}}%
\\ 
-%
{\displaystyle{p_z^2 \over \hbar m\Omega _z\coth (\beta \hbar \Omega _z/2)}}%
\end{array}
\right\} .  \label{vclas}
\end{equation}
If moreover the parabolic confinement were neglected, which corresponds to $%
\hbar \Omega _{xy},\hbar \Omega _z\ll 1/\beta $ in (\ref{vclas}), one
finally would be left with the Maxwell-Boltzmann momentum distribution for
free particles: 
\begin{equation}
n_{\text{maxwell}}\left( {\bf p}\right) =\left( \frac \beta {4\pi m}\right)
^{3/2}\exp \left\{ -\beta \frac{p_x^2+p_y^2+p_z^2}{2m}\right\} .  \label{max}
\end{equation}

In experiments on Bose-Einstein condensed atomic vapors (e.g. \cite{jila2}),
the temperature of a cloud of parabolically trapped bosons was estimated by
fitting the {\sl Maxwell-Boltzmann distribution}{\it \ }(\ref{max}) to the
tail of the experimentally measured momentum distribution $n_{\text{measured}%
}{\bf (p})$. This can be done by first choosing a sufficiently large
threshold momentum $p_c$ such that atoms with a momentum larger than $p_c$
do not belong to the Bose-Einstein condensate, and\ then minimizing $%
\int_{p>p_c}\left( n_{\text{maxwell}}\left( {\bf p}\right) -n_{\text{measured%
}}\left( {\bf p}\right) \right) ^2d{\bf p}$ with respect to the temperature $%
T$ appearing in the Maxwell-Boltzmann distribution. The temperature obtained
by this minimization will be denoted as $T_{\text{MB}}$ and the procedure
will be referred to as the ``Maxwell-Boltzmann fit procedure''.

Since the Maxwell-Boltzmann momentum distribution is a rather crude
approximation to the momentum distribution (\ref{vid}) of a gas of
non-interacting, parabolically trapped bosons, we rather propose to
determine the temperature by fitting $n_{\text{experiment}}\left( {\bf p}%
\right) $ to the momentum distribution (\ref{vid}) derived in the present
paper. This is done by minimizing $\int_{\text{all {\bf p}}}\left( n_{\text{%
bose}}^{\text{ideal}}\left( {\bf p}\right) -n_{\text{measured}}\left( {\bf p}%
\right) \right) ^2d{\bf p}$ with respect to the temperature $T$ appearing in
the momentum distribution $n_{\text{bose}}^{\text{ideal}}\left( {\bf p}%
\right) $ for non-interacting, parabolically trapped bosons. We will denote
this procedure as the ``Bose-Einstein fit procedure'' and the resulting
temperature as $T_{\text{BE}}$.

In order to compare the Maxwell-Boltzmann fit procedure with the
Bose-Einstein fit procedure, we applied the former on a momentum
distribution simulated with $n_{\text{bose}}^{\text{ideal}}\left( {\bf p}%
\right) $ ($\approx n_{\text{measured}}\left( {\bf p}\right) $ for weakly
interacting bosons), at a given temperature $T_{\text{BE}}$. That is, $%
\int_{p>p_c}\left( n_{\text{maxwell}}\left( {\bf p,}T_{\text{MB}}\right) -n_{%
\text{bose}}^{\text{ideal}}\left( {\bf p},T_{\text{BE}}\right) \right) ^2d%
{\bf p}$ was minimized with respect to $T_{\text{MB}}$, keeping $T_{\text{BE}%
}$ and $p_c$ fixed.

The result $T_{\text{MB}}$ of the minimization is compared to $T_{\text{BE}}$
in Figure 2, where the ratio $(T_{\text{BE}}-T_{\text{MB}})/T_{\text{BE}}$
is shown as a function of $T_{\text{BE}}$ at a fixed threshold momentum $%
p_c=5\sqrt{\hbar mw}$. In the Bose-Einstein condensed phase the discrepancy
between $T_{\text{BE}}$ and $T_{\text{MB}}$ increases when the number of
bosons increases, whereas for $T_{\text{BE}}>T_0=\hbar w\sqrt[3]{N/\zeta
\left( 3\right) },$ the discrepancy decreases when the number of bosons is
augmented, as shown in the inset of figure 2. For temperatures well above
the condensation temperature $T_0$, the ratio $(T_{\text{BE}}-T_{\text{MB}%
})/T_{\text{BE}}$ is small, and the Maxwell-Boltzmann fit procedure gives a
reliable temperature estimate in this high-temperature regime. The
difference between $T_{\text{BE}}$ and $T_{\text{MB}}$ becomes appreciable
below the condensation temperature $T_0$. The present analysis therefore
shows that $(T_{\text{BE}}-T_{\text{MB}})/T_{\text{BE}}$ is not negligible
in the Bose-Einstein condensed phase and that the Bose-Einstein fit
procedure is more appropriate than the Maxwell-Boltzmann fit procedure for
the determination of the temperature.

\section{Conclusion}

In this paper, we calculated the momentum distribution function (\ref{exact}%
) for a Bose gas in an anisotropic parabolic trapping potential, including
quadratic interparticle interactions. In the absence of interparticle
interactions, this momentum distribution simplifies to the momentum
distribution function (\ref{vid}) of an ideal boson gas in a harmonic
confining potential. The Maxwell-Boltzmann distribution (\ref{max}) follows
from (\ref{vid}) by the further simplifications of neglecting the boson
statistics and the parabolic confinement. The underlying assumption in these
derivations is that the number of bosons is sufficiently large to maintain
the temperature concept from the thermodynamical limit. Assuming furthermore
thermodynamical equilibrium, two temperature estimates were compared. Since
the temperature estimates used in experiments on Bose-Einstein condensed
atomic vapors (e.g. \cite{jila2}), are obtained by fitting the
Maxwell-Boltzmann distribution{\it \ }(\ref{max}) to the tail of the
experimentally measured momentum distribution $n_{\text{measured}}{\bf (p}),$
we derived the corresponding temperature $T_{\text{MB}}$ by fitting the
large-momentum tail of the Maxwell-Boltzmann distribution to the
large-momentum tail of the momentum distribution of the Bose gas (\ref{vid}%
). This temperature estimate was compared to the temperature $T_{\text{BE}}$
in the momentum distribution of the boson gas. We find that $T_{\text{MB}}$
substantially deviates from the correct temperature $T_{\text{BE}}$ in the
Bose-Einstein condensed phase, and hence that the Bose-Einstein fit
procedure is more appropriate than the Maxwell-Boltzmann fit procedure for
the determination of the temperature.

\section*{Acknowledgments}

Part of this work is performed in the framework of the FWO\ projects No.
1.5.729.94, 1.5.545.98, G.0287.95, G.0071.98, and WO.073.94N
(Wetenschappelijke Onderzoeksgemeenschap over ``Laagdimensionele systemen'',
Scientific Research Community of the FWO on ``Low Dimensional Systems''),
the ``Interuniversitaire Attractiepolen -- Belgische Staat, Diensten van de
Eerste Minister -- Wetenschappelijke, Technische en Culturele
aangelegenheden'', and in the framework of the BOF\ NOI 1997 projects of the
Universiteit Antwerpen. Two of the autors (J.T. and F.B.) acknowledge the
FWO for financial support.

\begin{center}
{\bf Figure Captions}
\end{center}

{\bf Fig.1:} In Figure 1 the average number $n({\bf p})/n({\bf p}=0)$ of
particles with a momentum ${\bf p},$ is shown at several temperatures for a
Bose gas in an isotropic parabolic confinement, as function of the momentum.
Units are chosen such that $\hbar =m=\Omega =1$. In this set of units, the
momentum is expressed in $p_0=\sqrt{\hbar m\Omega }$. In the inset, the
momentum distribution at ${\bf p}=0$ is shown as a function of temperature,
with $T_0=\hbar \Omega /k_B\sqrt[3]{N/\zeta (3)}$. For this figure we have
chosen $w=\Omega $ : the momentum distribution for $w/\Omega \neq 1$ is
qualitatively similar to the distribution for $w/\Omega =1$.

{\bf Fig.2:} The relative difference between the temperature $T_{\text{MB}}$
found by fitting a Maxwell-Boltzmann distribution and the actual temperature 
$T_{\text{BE}}$ is shown in this figure, representing $(T_{\text{BE}}-T_{%
\text{MB}})/T_{\text{BE}}$ versus $T_{\text{BE}}$ in units $T_0=(\hbar
\Omega /k_B)\sqrt[3]{N/\zeta (3)}$. This is shown for 500, 1000 and 2000
non-interacting bosons. In the inset, the dependence of $(T_{\text{BE}}-T_{%
\text{MB}})/T_{\text{BE}}$ on the number of particles is shown.

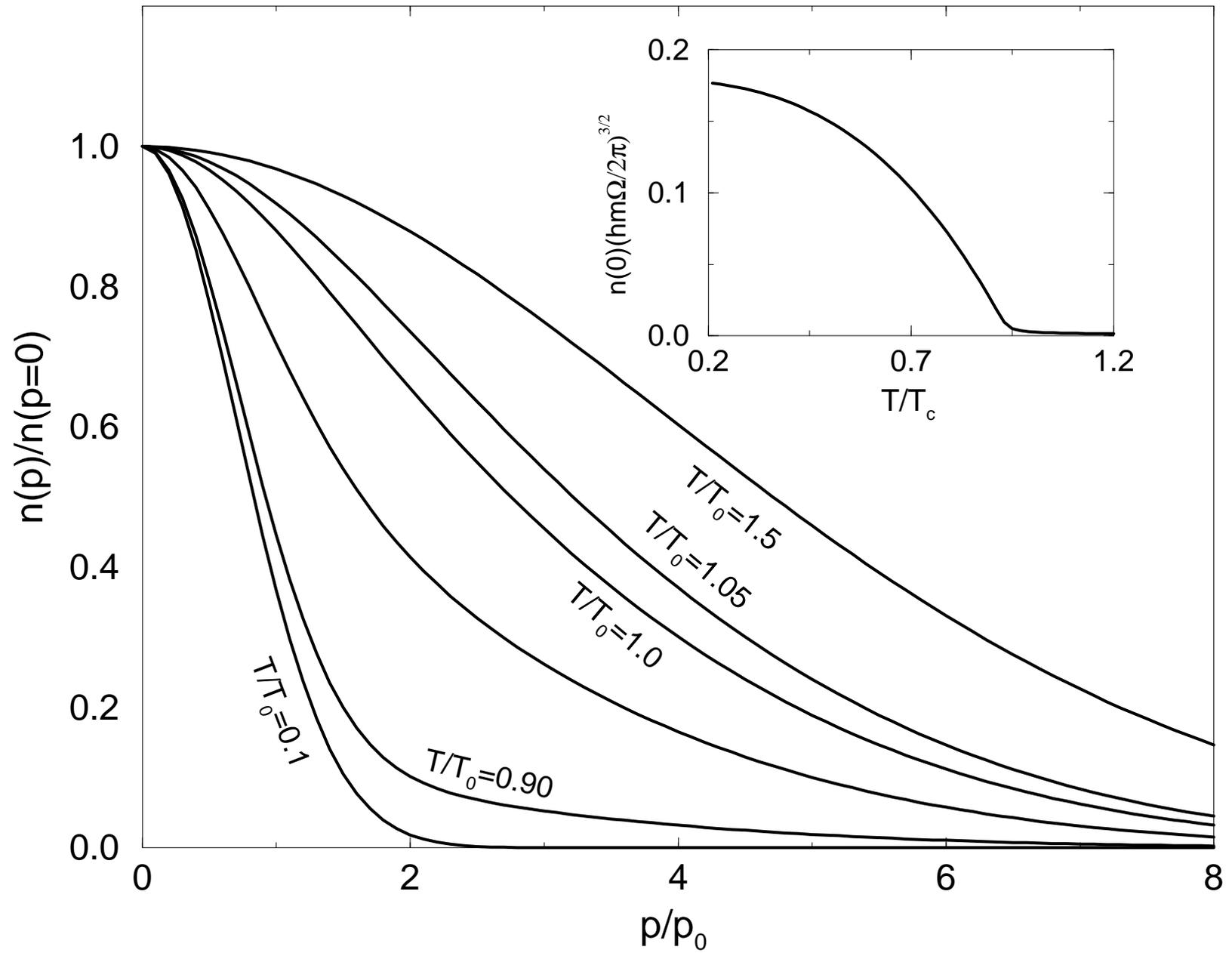
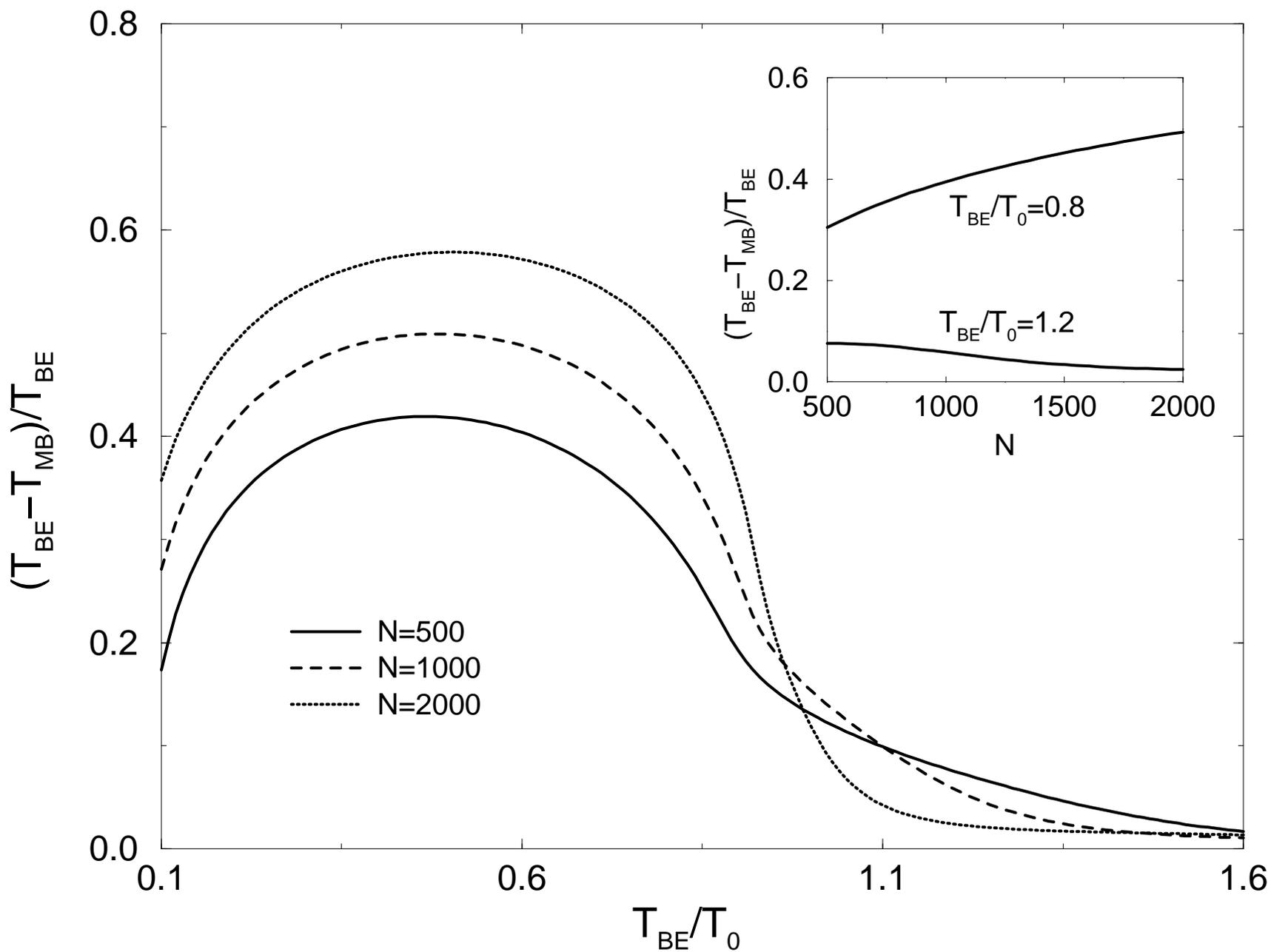
\end{document}